\def\G1915{GRS $1915$+$105$}
\def\X1550{XTE J$1550$--$564$}
\def\J1655{GRO J$1655$--$40$}

\def\etal{et al. }

\documentclass[]{aastex}
\usepackage{emulateapj5}
\usepackage{epsfig}
\usepackage{epsf}
\usepackage{color}
\definecolor{red}{rgb}{0.7,0,0}
\definecolor{blue}{rgb}{0,0,0.7}

\submitted{Accepted for publication in ApJ}
\shorttitle{\X1550 Outburst 2000 : Timing Analysis}
\shortauthors{Rodriguez \etal}
\begin{document}
\title{An X-ray Timing Study of XTE J1550--564: Evolution of the Low Frequency QPO for the Complete 2000 Outburst}

\author{J. Rodriguez\altaffilmark{1,2}, S. Corbel\altaffilmark{3,1}, E. Kalemci \altaffilmark{4}, 
J.A. Tomsick\altaffilmark{5} and M. Tagger\altaffilmark{1}}

\altaffiltext{1}{DSM/DAPNIA/Service d'Astrophysique (CNRS FRE 2591), CEA Saclay, 91191 Gif sur Yvette, France}
\altaffiltext{2}{ISDC, Chemin d'Ecogia, 16, 1290 Versoix, Switzerland}
\altaffiltext{3}{Universit\'e Paris VII Denis Diderot, 2 place Jussieu, 75005 Paris, France.}
\altaffiltext{4}{Space Sciences Laboratory, University of California, Berkeley, CA 94702-7450, USA}
\altaffiltext{5}{Center for Astrophysics and Space Science, Code 0424, University of California 
at San Diego, La Jolla, CA, 92093-0424, USA}

\begin{abstract}
We report on {\em RXTE} observations of the microquasar \X1550 during a 
$\sim 70$ day outburst in April-June 2000.  We focus here on the temporal 
properties of the source and study the behavior of low frequency ($0.1-10$ 
Hz) quasi periodic oscillations (LFQPO), which seem to be of  different 
types. We focus on the so-called type C (according to the classification 
of Remillard and collaborators), which corresponds to a strong $0.1-6$ Hz 
LFQPO, found to be present during at least 17 observations. 
We find that the frequency of the QPO is better correlated with the soft 
X-ray ($\leq 7$ keV) flux than with the hard flux ($\geq 7$ keV). 
If soft X-rays represent the behavior of an accretion disk, the relation 
shows that the disk may set the LFQPO frequency. In two cases, 
the identification of the type of QPO is not straightforward.  If the QPOs 
in those two cases are type A (or B), then we may be seeing the QPO type 
alternate between type C and type A (or B), and this may represent some 
rapid changes in the physical properties of the accretion flow, before the 
system stabilizes and slowly decays toward the end of the outburst.  On 
the other hand, if all the QPOs are of type C, we may be observing an 
inversion in the frequency vs. flux relation, similar to that seen in 
GRO~J1655$-$40. We discuss the QPO 
behavior in the framework of theoretical models.
 
\end{abstract}


\section{Introduction}
Soft X-ray transients (SXT) are accretion-powered binary systems, 
hosting a compact object (either a neutron star or a black hole), which 
spend most of their lives in quiescence, and are detected in X-rays as 
they undergo episodes of outburst. 
During an outburst, a given source can transit through spectral
 states defined by its spectral and temporal properties (see e.g.
McClintock \& Remillard 2003 for a recent review on spectral states).  
The power density spectrum  in the low/hard state 
(hereafter LHS) is characterized by strong aperiodic noise. The total 
fractional variability is high (often up to 50-60$\%$ rms), and strong 
($\sim 15\%$ rms) low frequency QPOs (0.1-10 Hz, hereafter LFQPO) are 
often observed.  The continuum of the power spectra can usually be approximated as a 
broken power-law with a flat power spectrum below $\sim 1$ Hz and a  
power spectrum that decreases with frequency above the break.
A powerful compact jet is an ubiquitous property of this state 
(Fender 2001). In the High/Soft State (also referred to as the Thermal 
Dominant state or TD), the fractional variability is weak (a few $\%$), 
and the power spectrum typically has a power-law shape, scaling roughly as 
the inverse of the frequency.  The power density spectra for the 
intermediate and very high states (IS/VHS) are similar to each other, 
with a fractional variability between $5$ and $\sim 20 \%$.  Low and 
high frequency QPOs are sometimes present in the IS/VHS (e.g. Remillard  
\etal 1999, and references therein; Homan \etal 2001; Miller \etal 2001). 
Based on the analysis of {\em Ginga} data, Rutledge et al. 
(1999) showed that the VHS is spectrally intermediate between the 
LHS and the TD. The similarity between VHS an IS may indicate that 
both are the same state observed at different luminosities (M\'endez 
\& van der Klis 1997, Homan et al. 2001). We will refer to this 
state as to Steep Power Law state (SPL, McClintock \& Remillard 2003). \\

\indent \X1550 was first detected by the All-Sky Monitor (ASM) on board 
{\em RXTE} on 1998 September 7 (Smith 1998) as it was transiting from 
quiescence into a LHS. It exhibited, a few days later, one of the brightest 
flares ($\sim 7$ Crab) observed with {\em RXTE}. Extensive spectral 
analysis is reported in Sobczak \etal (1999) and Sobczak \etal (2000a). 
Homan \etal (2001), based on the spectro-temporal properties of 
XTE J1550--564, showed that although the source was generally found in a 
TD during the second part of the outburst, hard flares and state 
transitions toward SPL and LHS could occur over a wide range of luminosity. 
They interpreted this behavior as evidence that spectral states are 
set not only by the mass accretion rate, $\dot{M}$, but that changes in 
at least one other physical parameter are also required.
Based on its spectral and temporal properties, it was believed 
early-on that \X1550 hosts a black hole rather than a neutron star 
(Sobczak \etal 1999, Cui \etal 1999).  Its black hole nature was confirmed 
with a compact object mass measurement of $M=10.5 \pm 1.0$ $ M_\odot$ 
(Orosz \etal 2002).  Low and high frequency QPOs have been detected in 
some {\em RXTE} observations (resp. $\sim 0.01-20$, and $102-285$ Hz), 
making \X1550 one of the eight black hole binaries for which high frequency 
QPOs have been reported (Remillard \etal 1999; Sobczak \etal 2000b; Homan 
\etal 2001, Miller \etal 2001, Remillard et al. 2002a, Homan et al. 2003a, 
Homan et al. 2003b).  Radio monitoring has shown jet features with an 
apparent motion $\ge 2c$ (Hannikainen \etal 2001), firmly establishing the 
source as a microquasar. \X1550 is also the first microquasar for which 
large scale X-ray jets have been observed (Corbel \etal 2002).  After a 
$\sim 9$ month outburst, \X1550 returned to quiescence in June 1999 
(MJD 51346).\\

\indent On 2000 April 6, \X1550 became active again (Smith \etal 2000)
after several months of quiescence, undergoing a new outburst that
lasted for $\sim 70$.  This outburst was monitored with {\em RXTE}, 
the {\em Unconventional Stellar Aspect (USA)} experiment, and {\em Chandra}. 
Reilly \etal (2001) report on the {\em USA} observations of the entire 
period, and Tomsick, Corbel \& Kaaret (2001, hereafter TCK01) report on
the energy spectra measured by {\em RXTE} and {\em Chandra} during the 
decay of the outburst.  Radio observations showed evidence for a compact 
jet on MJD 51697, and a discrete ejection, possibly associated with 
a state transition (Corbel \etal 2001). Based on an analysis of 
the spectral evolution of \X1550 over its whole 2000 outburst using the 
{\em RXTE} data, we showed that the source transited from an initial LHS 
toward a SPL, which probably reflected a change in the relative importance 
of the emitting media, rather than an increase of the accretion rate 
(Rodriguez, Corbel \& Tomsick 2003, hereafter RCT03). In addition, we 
proposed that the decrease of the X-ray luminosity right after the maximum 
peak on MJD 51662 might be due to the ejection of coronal material. It is 
also interesting to note that the cut-off in the energy spectrum disappears 
at the same time (RCT03).  During the SPL, the low values of the color 
radius returned from the fits and their steadiness over $\sim 10$ days 
suggest that the inner disk is close to its Last Stable Orbit (LSO), in 
good agreement with the detection of a HFQPO with little variation in
frequency over this time period (Miller \etal 2001).   Kalemci \etal (2001) 
(hereafter K01) studied the temporal behavior of the source as it was 
entering the LHS, and they report on the detection of a $65$ Hz QPO. 
K01 discuss the possibility that this QPO could be of the same type as the 
higher frequency ($\geq 100$ Hz) QPO detected earlier during this outburst 
(Miller \etal 2001) at a frequency 249--276 Hz and suggest that if the QPO 
is related to the accretion disk (orbiting clumps, Lense-Thirring 
precession or diskoseismic oscillations), the frequency of the QPO may 
indicate that the inner edge of the disk moves outward during the decay of 
the outburst. This interpretation is compatible with the spectral 
studies of TCK01 and RCT03.\\

\indent We present here a detailed analysis of the evolution of a LFQPO 
during the 2000 outburst of \X1550, from MJD 51644 until MJD 51690. To 
the previous study of K01 (covering MJD 51680--51690), we add the complete 
analysis of the LFQPO during the initial (MJD 51644--51678) part of the 
outburst, including the beginning of the transition to the LHS. 
The organization of the paper is as follows:  We start by presenting the 
data reduction and analysis methods used.  Then, we describe the 
characteristics and evolution of the LFQPOs.  Finally, we discuss our
results in the framework of theoretical models.

\section{Observations and Data Reduction}
\label{sec:analysis}
We reduced and analyzed the {\em RXTE} data from the dates preceding 
those presented in K01, i.e. from April 12 (MJD 51646) to May 14 (MJD 51678), 
and we incorporate the results from K01 below, allowing us to study the complete outburst.  
We used the {\itshape {LHEASOFT}} package v5.3 for our analysis.  
We restricted ourselves to the times when the target elevation angle was 
more than $10^{\circ}$ above the limb of the Earth, the satellite
pointing was within $0.02^{\circ }$ of the target, and we rejected the 
data taken while crossing the South Atlantic Anomaly. 
We then extracted Proportional Counter Array (PCA) light curves using
all Proportional Counter Units (PCUs) that were simultaneously 
turned on over a single observation.  The data we used came from 
several different {\em RXTE} programs (P50134, P50137, and P50135), 
and not all the observations used the same PCA modes, leading to 
different data formats.  However, in every case, we were able to 
extract light curves with $2^{-7}$ s ($\sim 8$ ms) time resolution.
For each observation, we extracted four light curves in the energy
ranges shown in Table \ref{tab:energychannel}, and we also extracted 
a 2--40 keV light curve.  We produced Poisson noise corrected power 
spectra for each observation using {\em{POWSPEC V1.0}} over intervals 
of 64 s duration, between $0.015$ Hz and $64$ Hz. We averaged power 
spectra of observations taken the same day and applied geometrical 
rebinning in all cases. For each power spectrum, we estimated the 
background rate using {\em{pcabackest}} v3.0, and used the average 
rate in order to estimate the true rms normalization following, e.g., 
Berger \& van der Klis (1994) and K01\footnote{Note that two 
observations were made after the loss of the propane layer in PCU 0. 
However, this unit was turned off during both observations so that 
no PCU 0 background estimate was required.}\\

\indent In order to characterize the QPO with the best possible 
sampling, we included in our study the data obtained with the {\em USA} 
satellite, as described in Reilly et al. (2001).  In particular, we 
focus on both the QPO centroid frequency and its fractional amplitude. 
Note that the latter is barely comparable with that obtained from the 
{\em RXTE} observations due to the fact that the {\em USA} energy 
range of 1--16 keV is lower than the {\em RXTE} energy range.
 

\begin{table*}[htbp]
\centering
\begin{tabular}{ccc}
Abs. Channels & Epoch 4 (until May 12, 2000) & Epoch 5\\
\hline
0--9 & $2-4.21$ keV & $2-4.09$ keV \\
10--17 & $4.21-7.56$ keV & $4.09-7.35$ keV\\
18--35 & $7.56-15.19$ keV & $7.35-14.76$ keV\\
36--91 & $15.19-39.81$ keV & $14.76-38.66$ keV \\
\hline
\end{tabular}
\caption{Absolute channels and the corresponding energy ranges over which are 
extracted the light curves}
\label{tab:energychannel}
\end{table*}

\begin{figure*}
\centering
\epsfig{file=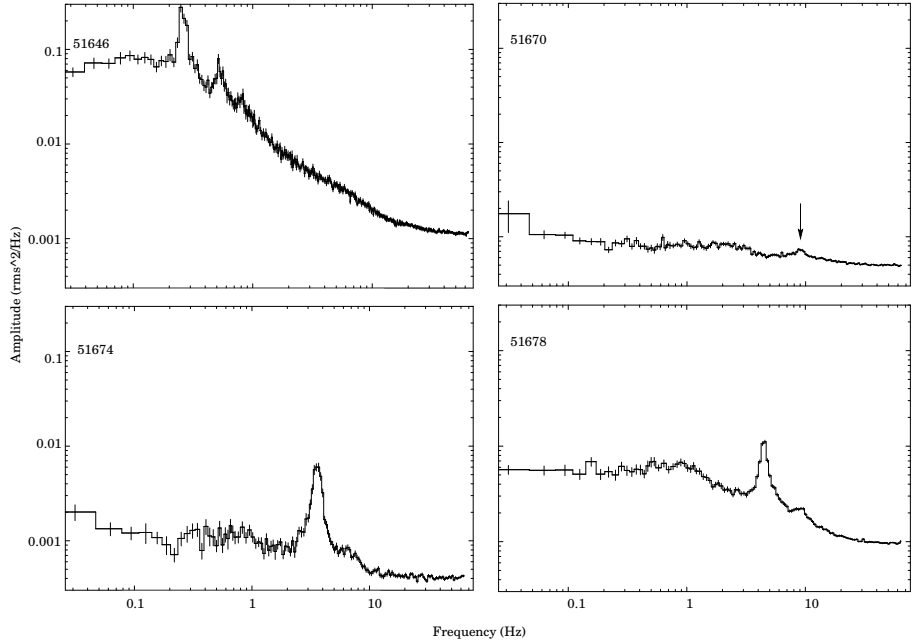,width=12cm} 
\caption{Typical power spectra of the source during the two spectral states 
studied in the course of our analysis. Power spectra extracted on MJD 51646 (LHS), MJD 51670 (SPL), MJD 51674 (SPL), and MJD 51678 (SPL). The graphs are plotted with the same vertical scale, for comparison. In all cases white noise is not subtracted on the 
plots. The arrow shows the faint type A/B QPO (see text for details). One can note the similarity between the three strong QPOs, in particular between MJD 51674 and MJD 51678.}
\label{fig:powspec}
\end{figure*}

\section{Results}
\subsection{Power Spectra}
\begin{figure*}
\centering
\epsfig{file=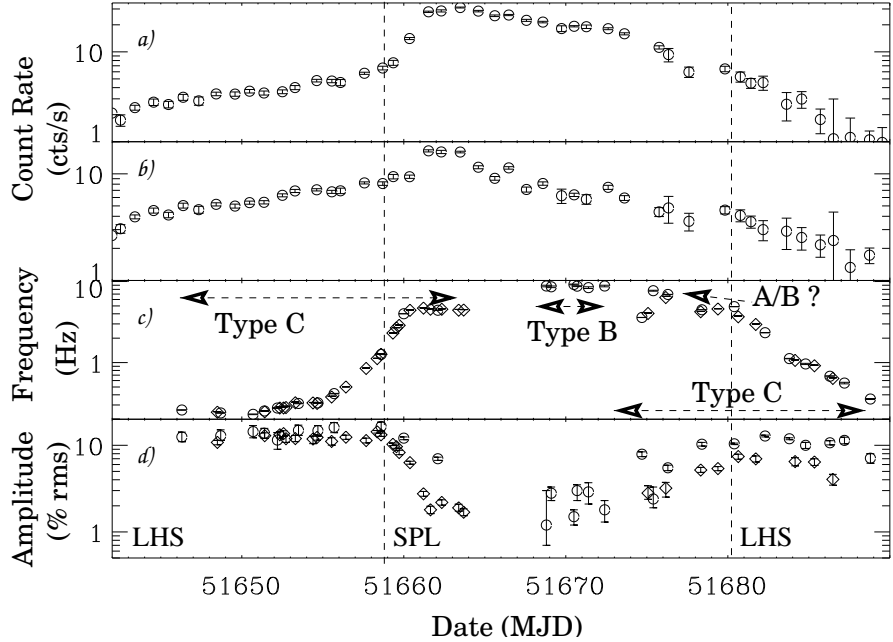,width=13cm}
\caption{ a) {\em RXTE}/ASM 1.2--3 keV and b) {\em RXTE} 5--12 keV  
light curves of XTE 1550--564 during the outburst.  Panel c) shows the evolution
of the frequency of the QPO detected with {\em RXTE}/PCA (circle, this study), and with the 
{\em Unconventional Stellar Aspect (USA)} (diamonds, Reilly et al. 2001). The arrows show
the type of the QPO. Panel d) shows the evolution of the QPO fractional amplitude with time. 
Diamonds are from Reilly et al. 2001 ({\em USA} data 1--16 keV), and circles from {\em RXTE} 
(2--40 keV).  The different energy ranges used in the analysis of the two data sets explain 
the difference of amplitude between both groups of points. The vertical dashed lines show the 
dates of states transitions (RCT03)}
\label{fig:overview2}
\end{figure*}

 We fit the 2--40 keV power spectra between 0.015~Hz and 20~Hz with a model 
consisting of Lorentzians. During the LHS, three such features are 
included in the modeling, whereas, during most of the SPL, two are sufficient 
to give satisfactory fits. In all cases, one of the Lorentzians is zero-centered, 
while the others represent ``bumps'' visible in the power spectra at higher 
frequencies (Fig. \ref{fig:powspec}).  Zero-centered Lorentzians are usually 
attributed to shot noise in the modeling of power spectra. However, recent 
studies have pointed out a more general use of this type of model in the cases 
of black hole binaries (Nowak \etal 2002; Belloni \etal 2002).  Indeed, the 
shape of a broad Lorentzian is similar to a broken power law: it is roughly 
flat below a break frequency  $\nu_{break}=\sqrt{\nu_{centroid}^2+(\Delta/2)^2}$, 
where $\Delta$ is the FWHM, and $\nu_{centroid}$ is the centroid frequency of 
the Lorentzian returned from the fits. Above $\nu_{break}$, the Lorentzian 
decreases roughly as a power-law.  In addition, some peaked noise is also 
detected in the $0.1-10$ Hz range in most of our power spectra.  These features 
are well fitted by narrow Lorentzians of Q values ($\nu_{centroid}/\Delta$) $\ge 2$, 
leading us to consider them as QPOs.  In addition, harmonically related QPOs are 
detected in some observations. The QPO parameters are reported in 
Table \ref{tab:QPO1fit}, while Fig. \ref{fig:overview2} shows the evolution of the 
frequency of the fundamental QPO during the outburst. In addition to the 
{\em RXTE} results, we superimposed the results obtained with the {\em USA} 
instrument (from Reilly et al. 2001). The points plotted after MJD 51678 are 
from K01.\\


\begin{table*}[htbp]
\centering
\begin{tabular}{c c c c c c c c}
\hline
DATE & QPO$_1$ Freq. & Q$_1$ & rms & QPO$_2$ Freq. & Q$_2$ & rms & $\chi^2$\\
(MJD) & (Hz) & & ($\%$) & (Hz) & & ($\%$) & (d.o.f.)\\
\hline
 51646.3 & $0.260 \pm 0.002$ & $6.0$ & $12.5_{-1.5}^{+1.7}$ & $0.530\pm0.007$ & $7.1$ & $6.4\pm1.4$ & 162.1 (150)\\
 51648.7 & $0.239 \pm 0.002$ & $9.2$ & $12.9_{-2.5}^{+2.2}$ & $0.48$ {\it (frozen)} & $>10$ & $< 5.7$ & 138.2 (113)\\
 51650.7 & $0.230 \pm 0.002$ & $5.7$ & $14.4_{-2.2}^{+2.5}$ & $0.46$ {\it (frozen)} & $>10$ & $< 5.6$ & 119.4 (112)\\
 51651.4 & $0.253 \pm 0.002$ & $5.4$ & $13.8_{-1.8}^{+1.9}$ & $0.505_{-0.008}^{+0.01}$ & $8.7$ & $4.8_{-1.6}^{+1.9}$ & 148.8 (110)\\
 51652.2 & $0.275 \pm 0.003$ & $7.9$ & $11.5\pm2.5$ & $0.59\pm0.1$  & $10.5$ & $4.5_{-1.8}^{+2.8}$ & 124.3 (110)\\
 51653.5 & $0.313 \pm 0.002$ & $6.0$ & $14.9_{-1.9}^{+2.2}$ & $0.64_{-0.02}^{+0.01}$ & $9.2$ & $4.6_{-2.1}^{+3.1}$ &  96.6 (110)\\
 51654.7 & $0.318 \pm 0.003$ & $5.8$ & $14.8_{-2.0}^{+2.1}$ & $0.63_{-0.01}^{+0.02}$ & $4.8$ & $7.7_{-1.7}^{+2.0}$ & 119.8 (110)\\
 51655.7 & $0.414 \pm 0.003$ & $7.0$ & $16.1_{-2.1}^{+2.2}$ & \multicolumn{3}{c}{Broad feature} & 145.5 (110)\\
 51658.6 & $1.261 \pm 0.006$ & $9.0$ & $16.5_{-1.9}^{+2.0}$ & $2.57 \pm 0.03$ & $10.7$ & $4.7_{-1.2}^{+1.3}$ & 145.6 (110)\\
 51660.0 & $4.002 \pm 0.005$ & $11.0$ & $12.1_{-0.5}^{+0.4}$ & $8.20 \pm 0.03$ &$7.7$ & $4.7\pm 0.3$ & 379.4 (280)\\
 51662.1 & $4.42 \pm 0.01$ & $7.0$ & $7.0 \pm 0.3$ & $8.88 \pm 0.05$ & $6.7$ & $2.5\pm0.2$ & 98.2 (85)\\
 51668.8 &$8.8_{-0.1}^{+0.2}$ & 8.2 & $1.2_{-0.5}^{+1.8}$ & & & & 122.4 (120)\\
 51669.1 &$8.6\pm0.2$ & 2.5 &$2.8\pm0.5$ & & & & 95.5 (78)\\
 51670.5 & $9.2 \pm 0.1$ & 7.5 &$1.5\pm 0.3$ & & & & 126.9 (94)\\
 51670.7 & $8.76 \pm 0.10$ & 2.87 &$3.0_{-0.7}^{+0.5}$ & & & & 128.1 (113)\\
 51671.4 &$8.4\pm0.2$ & 2.27 & $2.9\pm0.8$ & & & & 102.5 (89)\\
 51672.4 & $8.8\pm0.2$ & 3.9 &$1.8\pm0.5$ & & & & 107.2 (89)\\
 51674.7 & $3.60 \pm 0.01$ & $6.2$ & $7.9\pm0.5$ & $6.7 \pm 0.1$ & $2.1$ & $4.0 \pm 0.4$ & 88.5 (92)\\
 51675.4 & $7.7 \pm0.1$ & $5.8$ & $2.4_{-0.5}^{+0.9}$ & $16.7 \pm 0.2$ & $5.2$ & $2.5_{-0.4}^{+0.5}$ & 124.1 (113)\\
51676.3 & $6.92 \pm 0.02$ & $11.0$ & $5.5\pm0.4$ & $14.0_{-0.2}^{+0.1}$ & $16.2$  & $1.8_{-0.7}^{+1.6} $ & 126.6 (108)\\
 51678.4 & $4.48 \pm 0.01$ & $6.8$ & $10.3 \pm 0.5$ & $9.36\pm0.08$ & $8.7$ & $3.0_{-0.5}^{+0.7}$ & 142.1 (113)\\ 
\hline
\end{tabular}
\caption{Best fit parameters for the low frequency  QPO, and its harmonic, when
 present. 
See the text for the details of the modeling. The dates after MJD 
51680 (included) are reported in K01. The frequency is the centroid frequency,
 and Q is the ratio (centroid/FWHM). Errors are given at the one-$\sigma$ level.
Upper limits are at the $90 \%$ level.}
\label{tab:QPO1fit}
\end{table*}

\indent From the very beginning of the outburst (in the LHS) until MJD 51662, the 
$2-40$ keV power spectra show a strong (up to $16 \%$ rms amplitude) LFQPO that 
is detected along with its first harmonic (up to $9 \%$ rms amplitude).
We note, in some cases, the presence of a second harmonic, but the feature is 
weak and is not detected in majority of the observations.\\

\indent During the first part of the decline (MJD 51664--51672), in the SPL, 
while the energy spectra show the presence of a thermal component with a 
constant inner disk temperature around 0.8 keV (from MJD 51664 to MJD 51674, 
RCT03), we detect a $8.5-9$ Hz QPO with a rather low rms amplitude ($\leq 3\%$) 
in some observations (Table \ref{tab:QPO1fit}).  These QPOs are not reported 
in the Reilly \etal (2001) paper on the {\em USA} observations, and non-detection 
of these weak QPOs by {\em USA} may be expected due to the smaller effective 
area of {\em USA} compared to {\em RXTE}.  As discussed in detail in 
Sec. 4.1, these QPOs are a different type from the strong $\sim 10\%$ rms
amplitude QPOs seen in the LHS.  In some cases, we detect a rather broad feature,
which we do not consider to be a QPO as it has a Q value $\leq$ 2. On MJD 51664, 
for example, just after the soft X-ray maximum, we detect a broad feature with a 
centroid frequency  $6.51 \pm 0.08$ Hz, and a Q value equal to 1.7.  Broad 
features can be due to the presence of a frequency evolving QPO (as is sometimes 
the case for \G1915, e.g. Rodriguez \etal 2002a).  We produced a dynamical power
spectrum for the observation on MJD 51664 to check for frequency evolution in
the components of the power spectrum, and we do not see any evidence 
for a variable feature. \\

\indent On MJD 51674, while the source remains in the SPL, and the spectral 
parameters show no particular variations, we detect a LFQPO with a rather 
low frequency (3.60 Hz) compared to the previous day (Table \ref{tab:QPO1fit})
\footnote{We note, however, that if the transitions to the LHS are defined through 
sharp changes in temporal properties as presented in Kalemci et al. (2004), then 
the source was already in the LHS on MJD 51674. For the sake of clarity, however, 
we will, in the course of this paper, keep the state transitions identified in RCT03. 
They are shown on Fig.\ref{fig:overview2}}. A harmonic is also detected 
on MJD 51674.  Both features (QPO+harmonic) are present in the next 12 
observations (MJDs 51675-51690, K01). 
From Fig. \ref{fig:overview2}, one can see that the frequencies increase during 
the rise of the outburst, whereas they decrease during the return to quiescence. \\

\indent The power of the fundamental is high during the initial 
LHS and decreases significantly after the state transition.  At the maximum 
of the soft flux on MJD 51662, the rms amplitude has dropped to 
$7.0\pm0.3 \%$ (Fig. \ref{fig:overview2}). During most of the SPL, the QPO 
is weak and sometimes no QPO is detected (see also Reilly \etal 2001). For a 
1 Hz FWHM QPO, the $3\sigma$ upper limit on the rms amplitude is around 
$0.5 \%$ during the SPL ($\sim 0.3\%$ on MJD 51664, and $\sim 0.6\%$ on 
MJD 51671). From MJD 51674 through the end of our study, the trend of the 
amplitude is rather complex. On MJD 51674, when the source was in the SPL
state, the rms amplitude is $7.9 \pm0.5 \%$ (see Table \ref{tab:QPO1fit}). 
Interestingly, on the following day (MJD 51675), although there is no
significant change in the energy spectrum, the QPO properties are much
different with a drop in the rms amplitude to $\sim2.4\%$ and an increase
in frequency from 3.6 Hz to 7.7 Hz.  After that day, the amplitude slowly 
increases, reaching a local maximum once the transition to the LHS has
completed on MJD 51682 (K01). The mean amplitude during that period is 
slightly lower than that of the initial LHS.  Given the uncertainties on 
the harmonic parameters, it is difficult to deduce a precise behavior in 
that case (Fig. \ref{fig:overview2} and Table \ref{tab:QPO1fit}).
\begin{figure*}
\centering
\epsfig{file=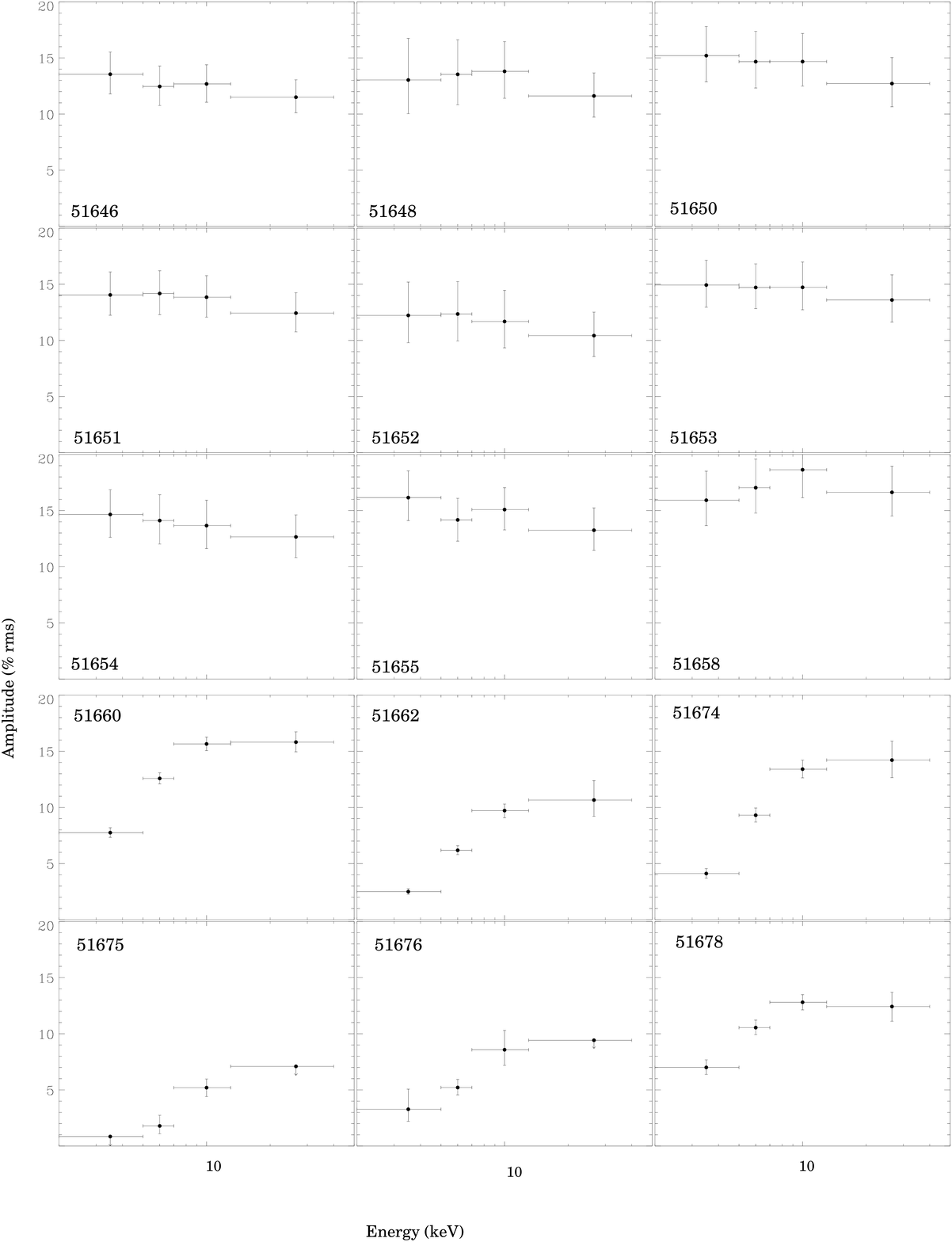,width=13cm}
\caption{Energy dependence of the LFQPO rms amplitude ($\%$) during the rise 
to outburst. 
The panels are time ordered from left to right, and from top to bottom. 
>From MJD 51646 through MJD 51658 the source is in the LHS (upper group), and 
later it is found in the SPL (lower group), until MJD 51680. The spectra after 
MJD 51678 are reported in Fig 4 a (SPL),b,c,d (LHS) of K01.}
\label{fig:QPO1spec}
\end{figure*}

\subsection{Spectro-temporal analysis of the LFQPO}
\label{sec:spectemp}
In order to study the energy dependent behavior of the QPO, we extracted 
power spectra in four energy bands as explained in section \ref{sec:analysis}. 
Each power spectrum is fitted with the same multi-Lorentzian model described
above. The resultant plots of QPO amplitude vs. energy (hereafter referred to 
as ``rms spectra'') for the fundamental are shown in Fig. \ref{fig:QPO1spec} 
(from MJD 51680 through MJD 51690 the results are plotted in Fig. 4 in K01).\\
\indent During the LHS (MJD 51646--51658), at the $1\sigma$ level, the rms 
spectra are roughly flat (Fig. \ref{fig:QPO1spec}). A clear change 
is observed during the SPL (MJD 51660--51678), where the rms spectra increase 
with energy, and seem to flatten in the highest energy band. From MJD 51664 
through MJD 51674, the $8.5-9$ Hz feature 
appears weak in the 2--40 keV band ($\leq 3 \%$). When we look at the energy 
dependence of its fractional amplitude, it is detected only in the 7--15 keV 
range, at a weak level ($\leq 5 \%$), and it is not visible in the other energy 
bands.\\

\subsection{Evolution of the Type C QPO frequency with the flux}
 Given the errors on the disk inner radius returned from the fits (RCT03), the 
frequency-radius, or frequency-temperature relation cannot be explored  
here.  Therefore, we first focus on the dependence of the QPO frequency with the 
flux, which also allows us to study the QPO behavior over the whole data set.\\

\begin{figure*}
\centering
\epsfig{file=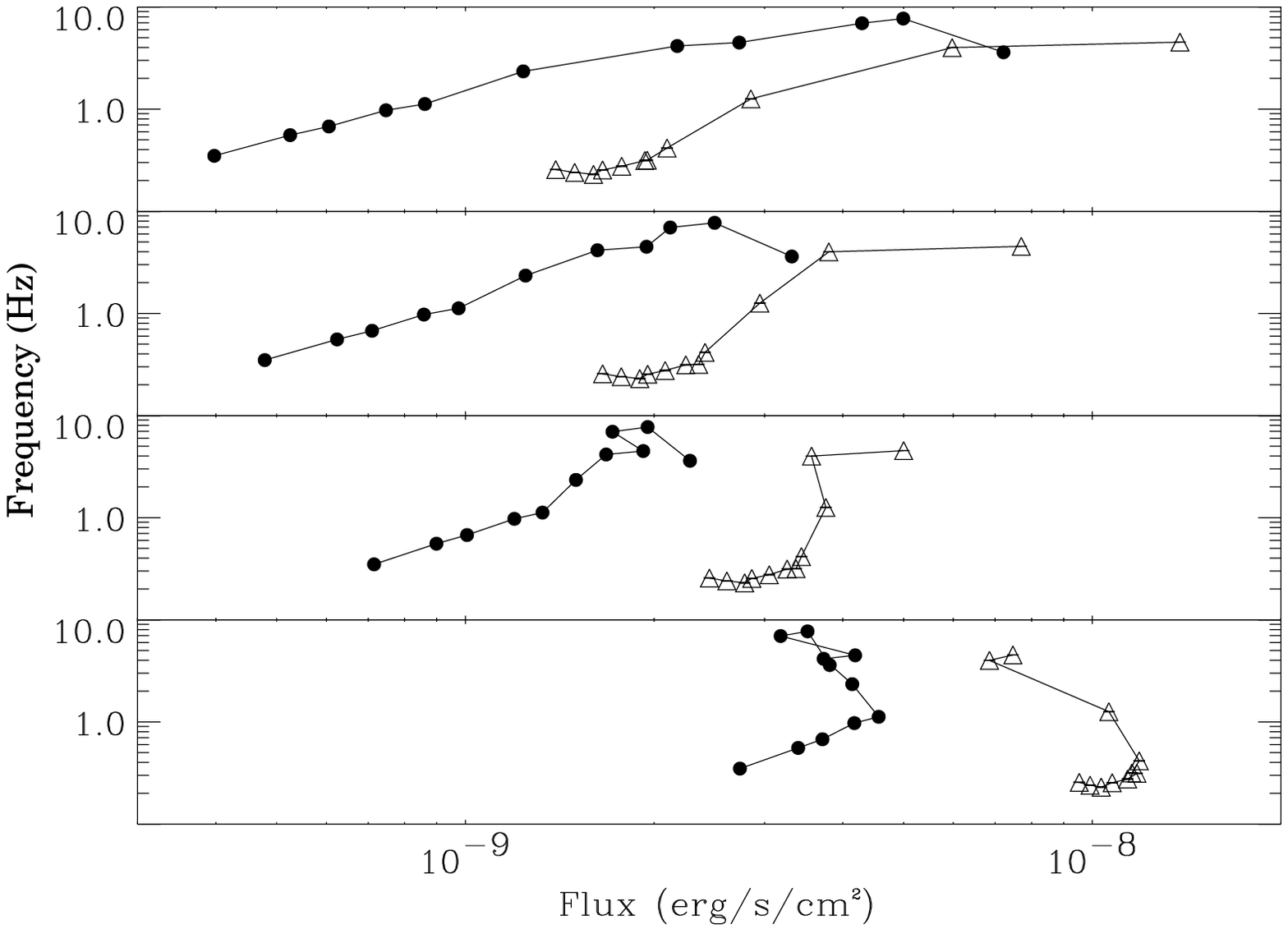,width=12cm}
\caption{Evolution of the QPO frequency with the flux in the four PCA bands
 defined in the text, {\it i.e.} from top to bottom $2-4$ keV, $4-7$ 
keV,  $7-15$ keV, and above $15$ keV. Triangles correspond to the rising 
part of the outburst (MJDs 51646--51662), and filled circles to the decay (MJDs 51674--51690).}
\label{fig:QPOvsFlux}
\end{figure*}

\indent The evolution of the QPO frequency with time, as seen by {\em RXTE} and 
{\em USA} is plotted in Fig. \ref{fig:overview2}, while the evolution of the 
frequency vs. the flux (as seen by {\em RXTE} only) in different energy bands 
is plotted on Fig. \ref{fig:QPOvsFlux}. When looking at the frequency 
evolution with time (Fig. \ref{fig:overview2}), it is quite interesting to note 
that the QPO appears/disappears at the same frequency during the rise and the 
decline of the outburst. After it reappears, the frequency increases and then 
decreases, while the ASM count rate decays. The possible inversion could have 
different origins as will be discussed in the next sections. 
For both epochs (initial LHS and final stage of the outburst from 51674 through 
51688), the evolution of the frequency correlates better with the $2-4$ and $4-7$ 
keV band fluxes. However, there are some differences between the two epochs. For 
the points corresponding to the rising part of the outburst, the evolution of 
the frequency vs. the 2-4 keV flux starts out flat and then becomes linear 
(above $\sim 1.6 \times 10^{-9}$ erg cm$^{-2}$ s$^{-1}$), up to 
$\sim 6 \times 10^{-9}$ erg cm$^{-2}$ s$^{-1}$, where a flattening (hereafter 
plateau) is visible. A similar trend is seen in the 4--7 keV band.\\
\indent For the points corresponding to the decline, the evolution of the 
frequency with the 2--4 keV flux is linear up to 
$\sim 5 \times 10^{-9}$ erg cm$^{-2}$ s$^{-1}$. Then, the frequency decreases 
from $7.7\pm0.1$ Hz to $3.60\pm0.01$ Hz, while the 2--4 keV flux is still 
increasing (Fig. \ref{fig:QPOvsFlux}). The 4--7 keV band shows a similar 
trend. We observe two opposite directions of the QPO frequency vs. flux 
evolution: a more common one, where the frequency is positively correlated 
with the (soft) flux, and an inversion of this correlation, although based on 
few points, occurring at higher flux. The plateau  
(Fig. \ref{fig:QPOvsFlux}) occurs at high soft flux 
($\geq 2\times10^{-9}$ erg cm$^{-2}$ s$^{-1}$) at times where the disk reaches 
the highest temperatures, and contributes the most to the energy spectra, 
meanwhile the power law contribution has decreased significantly (Fig. 
\ref{fig:overview2} and RCT03).  There are no such clear trends in the higher 
energy ranges (Fig. \ref{fig:QPOvsFlux}).\\

\section{Discussion}
\subsection{Two Different types of LFQPOs ?}
\label{sec:2QPOS}
The LFQPOs in \X1550 can be classified in three types (Remillard \etal 2002a, 
and references therein). The first type, type A, corresponds to broad QPOs 
(Q$\sim2-3$) with low coherence and negative time lags. The second ones 
(type B) are narrower (Q$\sim 10$), have high coherence, and positive 
time lags. Both these types have a rather faint fractional amplitude (a few $\%$), 
centroid frequencies between  $\sim6-10$ Hz, and are usually observed simultaneously 
with HFQPOs. Type C corresponds to narrow (Q $\geq$ 10) QPOs with variable frequency. 
They have a high coherence and usually display negative time lags, although 
positive time lags are sometimes seen with values up to $\sim 0.05$~s. 
They are usually not simultaneous  with HFQPOs. Fig. \ref{fig:powspec} shows 2--65 keV 
power spectra extracted  on MJD 51646 (strong LHS LFQPO), MJD 51670 (weak SPL QPO), 
and on MJD 51674 (strong SPL LFQPO). The strong QPOs we observe here, are likely 
to be of type C (see also the discussion in Reilly \etal 2001).\\
\indent The $\sim8.5-9$ Hz faint QPO appears at times where the soft bands are 
dominated by the thermal component (RCT03). Therefore, the non-detection of the QPO 
in the soft bands could simply reflect the fact that the disk photons are un-modulated, 
making the QPO appear fractionally weaker. However, the type C LFQPO has an amplitude 
that is usually higher than $5\%$ rms, whereas the $\sim8.5-9$ Hz QPO has an amplitude 
that is lower (3$\%$ at maximum). Even in the 7--15 keV range, where the $\sim8.5-9$ Hz 
QPO has its maximum rms amplitude, it is always weaker than the type C QPOs. 
This suggests that the QPOs are different types. The lack of (detected) harmonics 
may further confirm the QPO difference, although they might not be detected due to 
low fractional amplitudes. \\

\indent The $\sim8.5-9$ Hz QPO is observed at the same time that HFQPOs are observed 
(Miller \etal 2001). Although no lag study was performed for the observations 
before MJD 51674, a comparison with the analysis of Remillard et al. (2002a) indicates 
that the  $\sim8.5-9$~Hz QPO may be type B.\\

\indent The QPO on MJD 51674 has a low frequency, a high amplitude, and is 
rather narrow. In addition it has a high coherence and a slightly positive time lag 
($\sim0.01$~s, see Fig 4.40 in Kalemci 2002). It is therefore very likely that this 
QPO is type C. From MJD 51675 to MJD 51676 the QPOs do not satisfy all the criteria in 
the definition of type C QPOs, and may not be of type B either. This leads us to suggest 
the possibility that the QPOs seen on MJD 51675 and 51676 are type A, although the 
type C QPO has re-appeared on MJD 51674. From MJD 51678 through the end of our study,
the QPOs manifest all properties of type C (see also Fig. 4.42 in Kalemci 2002). 
 In the following, we will not consider type B QPOs further.  Furthermore, the 
non-detection of this feature in the lowest energy bands prevent any spectral study 
of this QPO as the one presented below.
 
\subsection{The Type C LFQPOs}
There have been several reports on a relation between QPO frequency 
and soft flux (e.g. GRS 1915+105, Markwardt, Swank \& Taam 1999; Rodriguez et al. 
2002b; 4U 1630--47, Trudolyubov et al. 2001). Such a correlation is, however, 
not always observed, as e.g. in XTE J1118+480 (Wood \etal 2000), or GX 339$-$4 
(Nowak, Wilms, Done 1999).  The correlation of the QPO frequency with the soft 
flux might indicate a connection with the Keplerian frequency of the inner boundary 
of the accretion disk. This interpretation seems in good agreement with 
our observations, especially during the outburst decline, during which the QPO frequency 
is linearly correlated with the soft flux (Fig. \ref{fig:QPOvsFlux}), which has been 
suggested to track changes in the accretion disk (RCT03, K01). The flatness of the 
frequency vs. Flux (Fig. \ref{fig:QPOvsFlux}) during the rise to outburst may appear 
in contradiction with such an interpretation. This apparent lack of correlation, 
however, most likely reflects the influence of the corona, and may be a signature of the  
source spectral hysteretic pattern (RCT 03, Kalemci 2002, Kalemci et al. 2004).\\

\indent The observation of a clear change of the shape of the rms spectra 
between  the LHS and the SPL (Fig. \ref{fig:QPO1spec}) is rather interesting. 
This change  is clearly related to the source state 
(Fig. \ref{fig:QPO1spec} in this work and Fig. 4 of K01).  At low energies, the 
amplitude of the QPO is anti-correlated with the soft X-ray flux,
which may simply reflects the influence of the disk on the overall 
spectrum of the source.  The local maximum at around 10 keV seen in 
K01, and the change in slope of the rms spectra (K01) at the end of the outburst 
raises challenging questions on the origin of the feature. The fact 
that the QPO rms amplitude is higher when the disk thermal component is weak 
or absent (and vice versa, RCT03) may indicate that the accretion disk is not 
the source of the modulated photons. 
We should note that a disk origin cannot completely be ruled out since some 
amplification by the corona could lead to the observed QPO spectra (e.g. Lehr 
et al. 2000). Furthermore, it is important to note that during the first LHS, 
the amplitude of the QPO seems to increase (although very little) when getting 
closer to the transition (at least until MJD 51658, Fig. \ref{fig:overview2}), 
while the evolution of the source flux is, in part, due to the approach of the 
accretion disk (RCT03).\\

\indent A connection of the QPO frequency with the Keplerian frequency, and an 
anti-correlation of its rms amplitude with the flux from the accretion disk can 
be understood, if the QPO is produced at 
a transition layer between the accretion disk and an inner hot accretion flow 
(Chakrabarti \& Titarchuk 1995). In that case, the QPO would represent radial 
oscillations of the transition layer (the CENBOL for CENtrifugal dominated 
BOundary Layer; Molteni, Sponholz \& Chakrabarti 1996; Chakrabarti \& Manickam 
2000), with little or no contribution from the disk photons to the QPO 
(Chakrabarti \& Manickam 2000).  The QPO frequency should be comparable to the 
free fall time scale in the inner hot region (Molteni \etal 1996). Therefore, 
a smaller inner disk radius will lead to a higher QPO frequency. \\

\indent Another possibility would involve a QPO produced by a ``hot spot,'' 
vortex, or spiral pattern rotating (at a characteristic frequency) in the disk as e.g. 
the Accretion Ejection Instability (AEI, Tagger \& Pellat 1999; Tagger et al. 
2004).  In this model, a spiral pattern rotates at $10-30 \%$ of the Keplerian frequency 
at the inner edge of the disk.  The relation between the QPO frequency and the 
flux is therefore a natural consequence of the instability. Note that in both of
these models, the disk photons do not contain the modulation, explaining the 
absence of connection between the disk luminosity and the QPO amplitude, in good
agreement with our observations.\\

\indent If the two QPOs observed on MJD 51675 and 51676 are of type A 
(or even B), as discussed in the previous section, it is interesting to note 
that the type C QPO re-appears on MJD 51674 at a frequency close to that 
observed just before its disappearance on MJD 51662 ($\sim 4$~Hz), and that it 
then shows the same kind of plateau value before it decreases after MJD 51680. 
This plateau frequency may indicate some saturation, or some limiting value in 
the mechanism(s) producing the type C QPO. We should note, however, that during 
the 1998-1999 outburst, type C QPOs were observed up to higher frequencies. The 
two outbursts are rather different, in particular in their durations and their
maximum luminosities.  It is, therefore, conceivable that the accretion flows 
manifest different characteristic size or time scales. It is worth mentioning 
that in either of the two aforementioned models (CENBOL and AEI), some kind of 
symmetry between the rise and fall is expected. We also note that changes in 
the physical properties of the source are seen between MJD 51662 and the 
following days (disappearance of thermal Comptonization, RCT03), and between 
51672 and 51674 (sudden increase of power-law flux, Kalemci et al. 2004). The 
transition between two different types of QPOs may indicate some rapid changes 
in the physical properties of the accretion flow, before the system stabilizes 
and slowly decays toward the end of the outburst. \\

\indent On the other hand, if the two QPOs observed on MJD 51675 and 
51676 are of type C, the inversion of the frequency-flux relation after 
MJD 51674 (Fig. \ref{fig:QPOvsFlux}), also visible in the {\em USA} data 
(Fig. \ref{fig:overview2}), is notable.  A clear inverted relation was seen
during observations of another black hole system, GRO~J1655$-$40 (Sobczak \etal 
2000b; Rodriguez \etal 2002b). In this system, the inverted relation was 
interpreted as being due to general relativistic effects when the disk was 
close to its last stable orbit (Varni\`ere et al. 2002). In the case of 
XTE~J1550$-$564, the reversal could have the same origin since it occurs at 
time when the accretion disk is close to its last stable orbit (RCT03, 
Kalemci et al. 2004).  This would be the first observation of a source 
showing the ``opposite'' behavior when the disk is close to the black hole, 
and the normal relation when it is further out.  However, the number of 
points involved in this inversion is small, and the behavior of the QPO 
frequency as a function of time has only been observed during this outburst. 
However, this does not necessarily mean that this is a unique case as a very 
frequent monitoring campaign was needed to catch the frequency inversion (or 
the rapidly changing QPO-type).  It will be interesting to see if similar 
behavior is observed from other SXTs with the ongoing observing programs of 
{\em RXTE}.

\begin{acknowledgements}
J.R. would like to thank Ph. Durouchoux, D. Barret, J-M. Hameury, G. 
Henry for careful reading of the manuscript and useful comments. J.R. also 
acknowledges financial support from the French Space Agency (CNES).
EK acknowledges partial support of T\"UB\.ITAK.
The authors warmly thank P. Ray for providing the {\em USA} data and the anonymous 
referee for useful comments which allowed to improve the quality of this 
paper.\\
This research has made use of data obtained through the High Energy 
Astrophysics Science Archive Center Online Service, provided by the NASA/
Goddard Space Flight Center.
\end{acknowledgements}

\end{document}